\documentclass[aps,prl,twocolumn,floatfix,superscriptaddress,nofootinbib,showpacs]{revtex4}

\usepackage{graphicx}     
\usepackage{dcolumn}      
\usepackage{bm}           
\usepackage{amsmath}
\usepackage{graphics}
\usepackage{amssymb}
\usepackage{amscd}
\usepackage{afterpage}
\usepackage{float,times}
\usepackage{epsfig}
\usepackage{theorem}
\usepackage{wrapfig}
\usepackage{picinpar}

\newcommand{\field}[1]{\mathbb{#1}}
 
\DeclareSymbolFont{AMSb}{U}{msb}{m}{n}
\DeclareMathSymbol{\N}{\mathbin}{AMSb}{"4E}
\DeclareMathSymbol{\Z}{\mathbin}{AMSb}{"5A}
\DeclareMathSymbol{\R}{\mathbin}{AMSb}{"52}
\DeclareMathSymbol{\Q}{\mathbin}{AMSb}{"51}
\DeclareMathSymbol{\I}{\mathbin}{AMSb}{"49}
\DeclareMathSymbol{\C}{\mathbin}{AMSb}{"43}

\def\R{\mathbb{R}}
\def\N{\mathbb{N}}
\def\C{\mathbb{C}}
\def\Z{\mathbb{Z}}

\begin{document}

\title{Scheme for direct measurement of a general two-qubit Hamiltonian.}

\author{Simon J. Devitt, Jared H. Cole, Lloyd C.L. Hollenberg}

\affiliation{
Centre for Quantum Computer Technology, School of Physics\\
University of Melbourne, Victoria 3010, Australia.}
\date{\today}

\begin{abstract}
The construction of two-qubit gates appropriate for universal quantum computation is of enormous importance 
to quantum information processing.  Building such gates is dependent on accurate knowledge of 
the interaction dynamics between two qubit systems.  This letter will present a systematic method for 
reconstructing the full two-qubit interaction Hamiltonian through experimental measures of concurrence.  
This not only gives a convenient method for constructing two qubit quantum gates, but can also be used to 
experimentally determine various Hamiltonian parameters in physical systems.  We show explicitly how this method 
can be employed to determine the first and second order spin-orbit corrections to the exchange coupling in quantum dots.
\end{abstract}


\maketitle

\begin{figure}[t]
\epsfig{figure=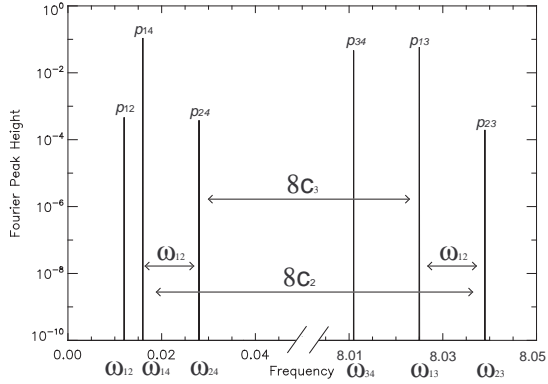,width = 0.9\linewidth}
 \caption{Power Density spectrum for the $|\psi_0\rangle_2$ input state.  
Note that the respective symmetries discussed earlier.  The values for each peak height are shown in table (\ref{tab:tab1}).}
 \label{fig:f1}
\end{figure}

\indent The DiVincenzo criterion for quantum computing \cite{DV} emphasises the importance of a universal set of 
gate operations for any physically achievable quantum computer.  Two qubit operations are an essential component 
of any universal gate library, creating entanglement between qubits.  This entanglement is what gives 
quantum computers extraordinary power over classical computational devices.  
The design of appropriate two qubit operations requires accurate 
knowledge of the interaction dynamics between qubits, specifically the two qubit interaction Hamiltonian.  
Accurate knowledge of this Hamiltonian is especially important for solid state devices, where qubit 
fabrication could easily lead to interaction dynamics which vary from qubit to qubit.
\\
\indent While in theory, many solid state systems rely on a purely Heisenberg type Hamiltonian \cite{he1,he2,he3}, anisotropic 
terms can be present when taking into account higher order effects.  For example, spin-orbit coupling in quantum 
dots \cite{dots1,dots2,dots3} where the Hamiltonian takes the time-\emph{in}dependant form,
\begin{equation}
H = J[S_1\cdot S_2 + \vec{\beta}\cdot (S_1 \times S_2) + S_1\Gamma S_2].
\label{equ:hamil}
\end{equation}
First order corrections are represented by $\vec{\beta}$, known as the Dzyaloshinski-Moriya (DM) vector \cite{DM1,DM2} 
and $\Gamma$ is a rank-3 symmetric tensor representing second order corrections.  Although theoretical 
estimates on these corrective terms have been made \cite{dots3}, and measurements of the DM corrections made in 
spin glass systems \cite{spin}, 
there has been no experimental measurement of either $\vec{\beta}$ or $\Gamma$ for isolated coupled dots.
\\
\indent In the context of quantum information processing, The system Hamiltonian is used to design quantum gate operations.  
A large amount 
of work has already been completed regarding the accurate construction of two qubit gates \cite{gate1,gate2,gate3,gate4,gate5}, 
this work can be split into two broad categories.  The first assumes that the quantum computing system exhibits a \emph{discrete} well defined 
two qubit interaction.  This occurs for architectures such as linear optics systems \cite{linear1,linear2}, with CNOT gates 
experimentally demonstrated \cite{cnot1,cnot2}.  These \emph{discrete} two qubit operations are traditionally characterised using state and 
process tomography \cite{cnot2,tom1,tom2,tom3} where the interaction Hamiltonian is initially assumed on well founded theoretical and/or 
experimental grounds.  Arbitrary gates can then be constructed directly from this fundamental discrete interaction \cite{gate4,gate5}.  While 
tomography has its advantages and is currently the method used by experimentalists, it cannot directly extract the Hamiltonian 
and becomes difficult when the gate is not ideal and a good theoretical model is unavailable.
\\
\indent The second category is exemplified by solid state architectures.  Here two qubit interactions 
exhibit time dependent control where 
unitary gates, $U(t)$, can be specified by the interaction Hamiltonian $H$, $U(t)=\text{exp}(iHt)$.  State and 
process tomography can be performed, but in general a \emph{discrete} gate has to be designed beforehand that can produce entangled states (for 
example a CNOT).  However, it is much more 
efficient to design interactions directly, using the
Hamiltonian \cite{gate1}.  
\\
\indent Methods for both single and two qubit Hamiltonian characterisation has recently been developed \cite{cole1,cole2,cole3}.  
These methods involve mapping the system evolution \cite{cole1,cole2} (single qubit characterisation) or the entanglement of the system \cite{cole3} 
(two qubit characterisation) over time.  
However, previous work on two qubit characterisation restricted the interaction Hamiltonian 
to a Heisenberg form, $H=c_1XX+c_2YY+c_3ZZ$, where $XX \equiv \sigma_x\otimes \sigma_x$, $YY \equiv \sigma_y\otimes \sigma_y$ and 
$ZZ \equiv \sigma_z\otimes \sigma_z$. 
In this letter, we present a more general method for characterisation.  We will show how mapping the entanglement of the 
two qubit system gives enough information to not only determine the entangling properties of a given interaction, 
but to fully reconstruct the Hamiltonian.  We apply this method to characterise a trial Hamiltonian of the 
form shown in Eq. \ref{equ:hamil} and therefore determine both $\vec{\beta}$ and $\Gamma$.
\\
\indent The most general form of a fully non local (FNL) two qubit Hamiltonian is 
$H=\sum_{i,j=1}^3d_{ij}\sigma_i \otimes \sigma_j$, where 
$\{\sigma_1,\sigma_2,\sigma_3\} = \{\sigma_x,\sigma_y,\sigma_z\}$ and $d_{ij} \in \field{R}$.  
Ref. \cite{gate1} shows that for any $d_{ij}$ there 
exists single qubit operators $K \in SU(2)\otimes SU(2)$ such that $U(t)=\text{exp}(iHt)=K^{\dagger}\text{exp}(iH_at)K$, where 
$H_a=c_1XX+c_2YY+c_3ZZ$ and $[c_1,c_2,c_3] \in \field{R}$.  
If $H_a$ and $K$ can be experimentally determined, then 
the original Hamiltonian can be reconstructed, $H=K^{\dagger}H_aK$.     
\\
\indent Determining $H_a$ and $K$ can be done by examining the entanglement produced by the operator $U(t)$ on a known \emph{separable} input state 
$|\psi\rangle$.  
The entanglement measure of the state $|\phi\rangle=U|\psi\rangle$, and hence $U$ can be quantified by the function $C^2 = |\langle\phi^*|YY|\phi\rangle|^2$, 
denoted the \emph{squared concurrence} \cite{ent1}. 
$C^2$ varies continuously between 0, for a product state and 1, for a maximally entangled Bell state.  Experimental techniques 
have been developed to measure this function for a given two qubit state that requires measurements in the $ZZ$ and $XZ$ basis \cite{ent3,cole3}.  
Our method for using $C^2$ to reconstruct the Hamiltonian will assume that single qubit gates have been characterised to arbitrary accuracy using the method detailed in \cite{cole2}, 
hence we have full single qubit rotation control and that measurements can be performed in these two bases.  
\\
\indent Consider an unknown Hamiltonian $H=\sum_{i=1}^3d_{ij}\sigma_i \otimes \sigma_j$.  $U(t)=\text{exp}(iHt)$ can be expressed according to 
\cite{gate1} as $U(t) = K^{\dagger}\text{exp}(iH_at)K$.  Given a \emph{known} separable input state $|\psi_0\rangle$, $C^2$ for the resultant  
state $U(t)|\psi_0\rangle$ can be written as $C^2 =
 |(Q^{\dagger}K|\psi_0\rangle)^TF^2(Q^{\dagger}K|\psi_0\rangle|^2$, where $F\equiv \text{diag}\{e^{i\lambda_1t},
e^{i\lambda_2t},e^{i\lambda_3t},e^{i\lambda_4t}\}$ and
$\{\lambda_1,\lambda_2,\lambda_3,\lambda_4\} = \{c_1-c_2+c_3,c_1+c_2-c_3,-c_1-c_2-c_3,-c_1+c_2+c_3\}$. 
The matrix $Q$ transforms 
computational states to Bell states \cite{gate1}.   
\\
\indent The added complexity in characterising FNL Hamiltonians compared with the non-isotropic Heisenberg Hamiltonian detailed in \cite{cole3} 
manifests itself in the $K$ matrix present in the expression for $C^2$.  In the Heisenberg case, $K=II$, where $II$ is the 4$\times$4 
identity matrix.  For FNL Hamiltonians, we have no knowledge of $H$, hence we have no knowledge of $K$.  The initial \emph{known} 
product state $|\psi_0\rangle$ is therefore rotated by $K$ before it is acted upon by the interaction term $\text{exp}(iH_at)$.  Unlike 
characterising Heisenberg Hamiltonians, $|\psi_0\rangle$ cannot be chosen to isolate $[c_1,c_2,c_3]$ directly.
\\
\indent Consider a general 4 $\times$ 4 matrix $K \in SU(2)\otimes SU(2)$.  It is known that the group $SU(2)\otimes SU(2)$ is 
isomorphic to $SO(4)$ \cite{gate1}, in fact $SO(4) = Q^{\dagger}SU(2)\otimes SU(2)Q$.  Hence $K_Q = Q^{\dagger}KQ \in SO(4)$ and 
all elements in $K_Q$ are real.  Let $Q^{\dagger}|\phi_0\rangle = K_QQ^{\dagger}|\psi_0\rangle = 
l_1|\Phi^+\rangle+l_2|\Phi^-\rangle+l_3|\Psi^+\rangle+l_4|\Psi^-\rangle$, be the rotated input product state in the Bell basis.  In this case, 
the concurrence can be rewritten as $C^2 = |e^{2i\lambda_1t}l_1^2+e^{2i\lambda_2t}l_2^2+e^{2i\lambda_3t}l_3^2+e^{2i\lambda_4t}l_4^2|^2$.
\\
\indent Experimentally, two qubits are initialised in a known product state $|\psi_0\rangle$ and $C^2$ mapped out as a function of $t$.  
Once $C^2$ has been mapped the data will have the following form,
\begin{equation}
\begin{aligned}
C^2 = \sum_{i=1}^4|l_i^2|^2 + \sum_{i,j\geq i}|l_i^2||l_j^2|e^{i(\omega_{ij}t+\gamma_{ij})},
\end{aligned}
\label{eq:e1}
\end{equation}
where $\omega_{ij} = 2(\lambda_i-\lambda_j)$ and $\gamma_{ij} = 2(\text{arg}(l_i)-\text{arg}(l_j))$.  The Fourier spectrum of the 
time series, $C^2$, is taken and the power density spectrum $|g(\omega)|^2$ is plotted as a function of $\omega$.
\\
\indent We will restrict this analysis to the case where $H$ decomposes to a form where $c_1\neq c_2\neq c_3\neq 0$.  This represents the 
most straightforward case for characterisation and arguably the most likely given a random Hamiltonian.  Other 
cases, i.e. when $c_1 = c_2 = c_3$, $c_3 = 0$ etc, can still be characterised, however the method is slightly more 
involved.  
The power density spectrum, $|g(\omega)|^2$, has the form,
\begin{equation}
\begin{aligned}
(2\pi)|g(\omega)|^2 = \bigg{(}\sum_{i=1}^4|l_i|^4\bigg{)}^2\delta(\omega) 
+ 
\sum_{i,j\geq i}|l_i|^4|l_j|^4\delta_{\omega_{ij}},
\end{aligned}
\label{eq:e2}
\end{equation}
with $\delta_{\omega_{ij}}=[\delta(\omega+\omega_{ij}) + \delta(\omega-\omega_{ij})]$. 
All the information on the rotated input state $K_QQ^{\dagger}|\psi_0\rangle$ is contained within the height of each peak, while 
all the information relating to the factors $[c_1,c_2,c_3]$ are related to the location of each peak in the frequency spectrum.
\\
\indent Extracting the actual matrix $K_Q$ and $[c_1,c_2,c_3]$ from the power density spectrum of Eq. \ref{eq:e2} is reasonably straightforward 
when $c_1\neq c_2\neq c_3\neq 0$.  Identifying each of the six peaks in the frequency spectrum relies on the principal of local equivalence \cite{gate1}.  
Decomposing any $H$ into $H_a$ defines a local equivalence class $[c_1,c_2,c_3]$ since any two operators $U_1$ and $U_2$ which 
are locally equivalent differ through single qubit rotations.  The values of $[c_1,c_2,c_3]$ are therefore locally periodic.  For example the 
single qubit operators $\pm iX\otimes X$ permutes $[c_1,c_2,c_3] \rightarrow [c_1 \pm \pi/2,c_2,c_3]$, and similarly for the local operators 
$\pm iY\otimes Y$ and $\pm iZ \otimes Z$.  This locally equivalent periodicity allows the restriction of $[c_1,c_2,c_3]$ such that $c_1 \geq c_2 \geq c_3 
\geq 0$.  Enforcing this condition allows the identification of all 6 separate peaks within the Fourier spectrum.  
\\
\indent From the definition of $\omega_{ij}$ we find, $\omega_{12} = 4(c_2-c_3)$, $\omega_{13}=4(c_1+c_3)$, $\omega_{14}=4(c_1-c_2)$, $\omega_{23}=4(c_1+c_2)$,
$\omega_{24}=4(c_1-c_3)$ and $\omega_{34}=4(c_2+c_3)$ (overall negative signs are omitted since Eq. \ref{eq:e2} is symmetric in $\omega_{ij} \rightarrow 
-\omega_{ij}$).  Using the restriction of $c_1 \geq c_2 \geq c_3 \geq 0$ and $c_1 \neq c_2 \neq c_3 \neq 0$, the two largest frequencies for all $c_i$ are 
$\omega_{23} > \omega_{13}$, which are separated by $\omega_{12}=4(c_2-c_3)$.  $\omega_{24} > \omega_{14}$ is also separated by $\omega_{12}$.  All 
four frequency peaks, $\omega_{23}$, $\omega_{13}$, $\omega_{14}$ and $\omega_{24}$ are cantered on the value $4c_1$.  The final peak is $\omega_{34}$, which 
with $\omega_{12}$ is symmetrically spaced about $4c_2 < 4c_1$, (see Fig. \ref{fig:f1}).
\\
\indent The coefficients $[c_1,c_2,c_3]$, which can be used to define $H_a$ can be determined, however to reconstruct $H$, $K$ such that $H = K^{\dagger}H_aK$ 
also needs to be found.  This can be done through the peak heights in the Fourier spectrum.  Consider the matrix $K_Q = Q^{\dagger}KQ \in SO(4)$.  All elements 
in $K_Q$ are real and six independent parameters are required to define an arbitrary $SO(4)$ matrix.  In the case where $c_1 \neq c_2 \neq c_3 \neq 0$ 
parameterisation of $K_Q$ in order to determine these six parameters isn't required.  Instead, all 16 elements of $K_Q$ can be found directly.  
Consider the following three separable states (neglecting normalisation), 
$|\psi_0\rangle_1 = (i|0\rangle+|1\rangle)\otimes (|0\rangle +i|1\rangle)$, 
$|\psi_0\rangle_2 = (|0\rangle+|1\rangle)\otimes |1\rangle$, 
$|\psi_0\rangle_3 = |1\rangle \otimes (|0\rangle + |1\rangle)$.
If we specify the matrix $K_Q$ through the 16 matrix elements $\{a_1,...,a_{16}\}$, then the squared modulus of the rotated Bell state coefficients, $|l_i^2(k)|$, 
for each of the three input states, $k$, are, $2|l_1^2(1)| = (a_1^2+a_3^2)$, $4|l_1^2(2)| = (a_1+a_3)^2+(a_2+a_4)^2$ and $4|l_1^2(3)|=(a_1+a_3)^2+(a_2-a_4)^2$.  
Similar expressions are obtained for $|l_2^2(k)|$, $|l_3^2(k)|$ and $|l_4^2(k)|$ using the appropriate rows from the matrix, $K_Q$.  
\\
\indent Examining Eq. \ref{eq:e2} for the power density spectrum of $C^2$, the height of each of the respective peaks are, 
$p_{ij}(k) = |l_i(k)|^4|l_j(k)|^4$.  Since $|l_i(k)| \geq 0$, we can determine the absolute value of the rotated input state (in the Bell basis) as,
\begin{equation*}
\begin{aligned}
|l_1| &= \bigg{(}\frac{p_{12}p_{13}}{p_{23}}\bigg{)}^{1/8} = \bigg{(}\frac{p_{12}p_{14}}{p_{24}}\bigg{)}^{1/8} = \bigg{(}\frac{p_{14}p_{13}}{p_{34}}\bigg{)}^{1/8},    
\end{aligned}
\end{equation*}
\begin{equation}
\begin{aligned}
|l_2| &= \bigg{(}\frac{p_{12}p_{23}}{p_{13}}\bigg{)}^{1/8} = \bigg{(}\frac{p_{12}p_{24}}{p_{14}}\bigg{)}^{1/8} = \bigg{(}\frac{p_{24}p_{23}}{p_{34}}\bigg{)}^{1/8}, \\
|l_3| &= \bigg{(}\frac{p_{13}p_{23}}{p_{12}}\bigg{)}^{1/8} = \bigg{(}\frac{p_{13}p_{34}}{p_{14}}\bigg{)}^{1/8} = \bigg{(}\frac{p_{23}p_{34}}{p_{24}}\bigg{)}^{1/8}, \\
|l_4| &= \bigg{(}\frac{p_{14}p_{24}}{p_{12}}\bigg{)}^{1/8} = \bigg{(}\frac{p_{24}p_{34}}{p_{23}}\bigg{)}^{1/8} = \bigg{(}\frac{p_{14}p_{34}}{p_{13}}\bigg{)}^{1/8}, 
\end{aligned}
\label{eq:e3}
\end{equation}   
dropping the index $k$.  Invoking the group conditions of $SO(4)$, specifically, that the 4 rows of $K_Q$ forms an orthonormal set, 
leads to the following,
\begin{equation}
\begin{aligned}
\mu_1(1) &= |l_1(1)|^2-|l_1(2)|^2+|l_1(3)|^2 = \frac{1}{2}(a_3-a_1)^2, \\
\mu_2(1) &= |l_1(1)|^2+|l_1(2)|^2-|l_1(3)|^2 = \frac{1}{2}(a_3+a_1)^2, \\
\mu_3(1) &= -|l_1(1)|^2+|l_1(2)|^2+|l_1(3)|^2 = \frac{1}{2}(a_2-a_4)^2, \\
\mu_4(1) &= 1-|l_1(1)|^2-|l_1(2)|^2-|l_1(3)|^2 = \frac{1}{2}(a_2+a_4)^2. 
\end{aligned}
\label{eq:e4}
\end{equation}
Similar equations can be used to determine the other three rows of $K_Q$.  Since $\mu$ represents the square of the matrix elements, 
there is a certain amount of ambiguity for each row of $K_Q$.  Eq. \ref{eq:e4} give rise to solutions in pairs, $(a_1,a_3)$ and $(a_2,a_4)$ which 
implies that various matrices can be formed for $K_Q$ which 
vary through permutations and/or sign flips of these pairs.  This ambiguity 
leads to $2^{16}$ matrices for $K_Q$.  
\\
\indent Noting that $K_Q \in SO(4)$ eliminates a large number of possibilities.  Each possible matrix satisfying Eq. \ref{eq:e4} can be 
numerically constructed and checked that $K_Q^TK_Q = II$ and $\text{Det}(K_Q) =1$.  However, even in the case where all the elements of $K_Q$ are distinct, there 
still exists a number of matrices satisfying the group conditions of $SO(4)$.  Each of these matrices, $K_Q$, combined with the already determined $H_a$, 
will lead to a set of locally equivalent Hamiltonians, $\{H_i\}$, i.e. Hamiltonians with the same values for $[c_1,c_2,c_3]$.  
\\
\indent In order to narrow down the possible Hamiltonians, we analytically construct all the operators, $\{U_i(t)=\text{exp}(iH_it)\}$ for each locally 
equivalent Hamiltonian.  For each $U_i(t)$, $C^2$ can be calculated for all three input states 
$|\psi_0\rangle_k$ at some arbitrary value of $t$.  Comparing the experimental data to the analytic calculations of $C^2$ will reduce the possible lists of 
Hamiltonians to a smaller subset $\{H_{i'}\}$ that have precisely the same analytical
form for $C^2$, for all input states $|\psi_0\rangle_k$. 
\\
\indent To isolate the correct Hamiltonian from this finite set, $\{H_{i'}\}$, analytically construct the function 
$C^2$ for some arbitrary input state $|\psi_0\rangle$ that produces \emph{distinct} functional forms (for $C^2$) for each $U_{i'}(t)$.  
Once the appropriate input state has been analytically determined, experimentally measure $C^2$ for some single arbitrary value of $t$.  The 
experimental and analytical value of $C^2$ can be compared for each $U_{i'}$ and the correct $H_{i'}$ determined.  
Once this procedure has been completed, only the matrices, $K_Q$, that produce the same target Hamiltonian will remain, satisfying all 
the experimental data. 
\\ 
\indent 
Although the characterisation process is essential to the efficient construction of arbitrary two qubit gates, it can also be used 
to experimentally determine other theoretical parameters governing a system.  
The following is a specific example to highlight the methodology.
Consider the exchange interaction between spin systems in quantum dots.  As mentioned in the introduction, spin-obit coupling 
introduces anisotropic corrections to the otherwise \emph{ideal} isotropic exchange Hamiltonian, $H=JS_1.S_2$.
When these correction are included, the time independent Hamiltonian 
describing the two qubit coupling takes the form in Eq. \ref{equ:hamil}.
Although gate design schemes have been developed to overcome these anisotropic terms \cite{dots2}, 
characterisation of this coupling is possible and allows us to determine both 
$\vec{\beta}$ and $\Gamma$ up to a factor of $J$.  
\\
\indent We will consider the trial Hamiltonian where $J = 1$, $\vec{\beta} = \{0.01,0.005,0.02\}$ and $\Gamma = \{0.003,
0.005,0.001,0.0015,0.0024,0.0009\}$ (representing the 6 independent parameters specifying a $3\times 3$, real, symmetric matrix).  
These values are motivated by the fact that $J=1$ leads to extracting $\vec{\beta}$ and $\Gamma$ exactly, 
$\vec{\beta}$ is chosen randomly but is consistent with the analysis of Kavokin \cite{dots3} for quantum dots in GaAs.
$\Gamma$ is roughly an order of magnitude smaller since it is a second order correction to $H$.  
\\
\indent Fig. \ref{fig:f1} shows the power density spectrum obtained using the input state $|\psi_0\rangle_2=(|0\rangle+|1\rangle)|1\rangle$.  
From the spectrum we can identify the relevant peaks as $\omega_{23}=8.039$, $\omega_{13}=8.025$, $\omega_{12} = 0.012$, $\omega_{24} = 
0.028$, $\omega_{14} = 0.016$ and $\omega_{34} = 8.011$.  These peak frequencies lead to 
$[c_1,c_2,c_3] = [1.006,1.003,0.999]$. 
The respective peak heights and values for ($\mu_1,\mu_2,\mu_3,\mu_4$) are given in table \ref{tab:tab1}. 
\\
\indent After constructing all possible combinations of the above values for $\{a_1,...,a_{16}\}$, 
64 matrices are found to satisfy the 
group conditions of $SO(4)$.  These matrices and $H_a$ are used to find 16 distinct Hamiltonians $H=K^{\dagger}H_aK$.  Eq. 
\ref{eq:ham} shows 4 Hamiltonians in terms of the spin-orbit corrections $\vec{\beta}$ and $\Gamma$.  
\begin{equation}
\begin{aligned}
H_1: & \quad \vec{\beta} = \{-0.0015,0.005,-0.009\} \\
 & \quad \Gamma = \{-2,0.005,-2,-0.01,0.002,-0.002\} \\
H_2: &\quad \vec{\beta} = \{0.01,0.005,0.02\} \\
&\quad \Gamma = \{0.003,0.005,0.001,0.0015,0.0024,0.0009\} \\
H_3: &\quad \vec{\beta} = \{0.009,0.005,0.0015\} \\
&\quad \Gamma = \{-2,0.005,-2,0.02,0.002,0.01\} \\
H_4: &\quad \vec{\beta} = \{0.02,0.005,0.01\} \\
&\quad \Gamma = \{0.001,0.005,0.003,0.0009,0.002,0.0015\} \\
\end{aligned}
\label{eq:ham}
\end{equation}
\begin{table}[t]
\begin{tabular}{|p{.9cm}|p{.9cm}|p{.9cm}|p{.9cm}|p{.9cm}|p{.9cm}|p{.9cm}|p{.9cm}|}
\hline Input & $p_{12}$ & $p_{13}$ & $p_{14}$ & $p_{23}$ & $p_{24}$ & $p_{34}$ & $\times 10^{-2}$\\
\hline $|\phi_0\rangle_1$ & 0 & 6.81 & 2.32 & 0 & 0 & 13.35& \\
\hline $|\phi_0\rangle_2$ & .0467 & 5.68 & 11.01& .0197 & .0383 & 4.66 & \\
\hline $|\phi_0\rangle_3$ & .0473 & 5.91 & 10.84& .0201 & .0367 & 4.59 & \\
\hline
\hline $\mu_1(1)$ & $\mu_2(1)$ & $\mu_3(1)$ & $\mu_4(1)$ & $\mu_1(2)$ & $\mu_2(2)$ & $\mu_3(2)$ & $\mu_4(2)$\\ 
\hline  21.21 & 20.46 & 56.01 & 2.33 & .0041 & .021 & 4.49 & 95.48\\
\hline 
\hline $\mu_1(3)$ & $\mu_2(3)$ & $\mu_3(3)$ & $\mu_4(3)$ & $\mu_1(4)$ & $\mu_2(4)$ & $\mu_3(4)$ & $\mu_4(4)$\\
\hline 50.24 & 49.74 & .0110  & .0012  & 28.55 & 29.78 & 39.49 & 2.19\\
\hline
\end{tabular}
\caption{Relevant data from the three power density spectra in Fig. \ref{fig:f1}.  Values for all the respective peak heights and 
$\{\mu_1(i),\mu_2(i),\mu_3(i),\mu_4(i)\}$ can be used to find $K_Q$.}
\label{tab:tab1}
\end{table}
When parameterising Eq. \ref{eq:ham} with respect to $\vec{\beta}$ and $\Gamma$, we have explicitly used 
$J=1$, this is \emph{NOT} a requirement to characterise the Hamiltonian using this method.
However, since the general form of the Hamiltonian has $J$ as a constant 
multiplicative factor, extracting the exact values for $\vec{\beta}$ and $\Gamma$ requires $J$ 
to be known or assumed.  
\\
\indent From the 16 possible Hamiltonians generated, clearly the $K_Q$ matrices generating $H_2$ are the 
target matrices, however we need to filter out the other 15 possible Hamiltonians.  To do this, analytically 
construct the unitary operators $U_i = \text{exp}(iH_i)$, for $i\in [1,16]$ and calculate $C^2$ 
for all $i \in [1,16]$ and $k \in [1,3]$.  The calculated values are then compared with the values of $C^2$ measured 
experimentally at $t=1$ for all three input states.  Table \ref{tab:final} shows the results for the 4 Hamiltonians 
in Eq. \ref{eq:ham}.
\begin{table}[b]
\begin{tabular}{|p{1.2cm}|p{1.2cm}|p{1.2cm}|p{1.2cm}|p{1.2cm}|p{1.4cm}|}
\hline Input & Ent($U_1$) & Ent($U_2$) & Ent($U_3$) & Ent($U_4$) & Ent($U(1)$) \\
\hline $|\phi_0\rangle_1$ & .581 & .581 & .581 & .581 & .581 \\
\hline $|\phi_0\rangle_2$ & .147 & .147 & .142 & .147 & .147 \\
\hline $|\phi_0\rangle_3$ & .144 & .150 & .150 & .150 & .150 \\
\hline
\end{tabular}
\caption{Calculations of $C^2$ for all three input states using the 4 unitary operators 
experimentally obtained for $t=1$.  The last column represents the data for the 
unknown unitary operator at $t=1$, $U(1)$.}
\label{tab:final}
\end{table}
Simulated values of $C^2$ immediately eliminate $H_1$ and $H_3$ as possible Hamiltonians.
Repeating for the other 12 Hamiltonians shows that the analytical values for 
$C^2$ match only for $H_2$ and $H_4$.  Here, the analytical forms for $C^2$ are identical for 
the three input states $|\psi_0\rangle_k$.  
\\
\indent To isolating $H_2$, we analytically construct  
$C^2$ and find that the input state $|\psi_0\rangle_4 = |1\rangle(|0\rangle-i|1\rangle)$ 
generates different functions for $H_2$ and $H_4$.  One more experimental measurement of $C^2$ is taken for this 
input state at $t=0.5$, giving $C^2 = 0.200$.  Analytically, $C^2(H_2) = 0.200$ and $C^2(H_4) = 0.196$.  
This final measurement has discriminated between $H_2$ and 
$H_4$, leaving $H_2$ to satisfy all the experimental measures of $C^2$.  
\\
\indent In conclusion, we have presented a systematic method to experimentally determine the interaction Hamiltonian 
of a two-qubit system that requires initialisation in a minimum of three product states and measurement in 
two separate bases.  We have demonstrated how this method not only allows for accurate 
characterisation of system Hamiltonians required for constructing two-qubit quantum gates, but also how characterisation 
can lead to the determination of other parameters of interacting systems.  Specifically, we showed how 
the first and second order anisotropic corrections, in quantum dots can be experimentally determined.  
A comprehensive error analysis of this scheme under a fixed number of measurements and how mis-characterisation 
propagates through to effective systematic gate errors is still required.  
\\
\indent The authors thank S. Schirmer and D. Oi for helpful discussions.  
This work was supported by the Australian Research Council, 
US National Security Agency, Advanced Research and Development Activity and Army Research Office under contract W911NF-04-1-0290.

\end{document}